\documentclass[12pt,preprint]{aastex}
\usepackage{color}

\shortauthors{Okamoto et al.}
\shorttitle{Helical motions of prominence threads}
\begin{document}

\title{Helical motions of fine-structure prominence threads observed by \emph{Hinode} and \emph{IRIS}}
\author{\textsc{
Takenori J. Okamoto,$^{1}$
Wei Liu,$^{2,3,4}$
Saku Tsuneta$^{5}$
}}
\altaffiltext{1}{National Astronomical Observatory of Japan, Mitaka, Tokyo 181-8588, Japan}
\altaffiltext{2}{Bay Area Environmental Research Institute, 625 2nd Street, Suite 209, Petaluma, CA 94952, USA}
\altaffiltext{3}{Lockheed Martin Solar and Astrophysics Laboratory, 3251 Hanover Street, Bldg.~252, Palo Alto, CA 94304, USA}
\altaffiltext{4}{W.~W.~Hansen Experimental Physics Laboratory, Stanford University, Stanford, CA 94305, USA}
\altaffiltext{5}{ISAS/JAXA, Sagamihara, Kanagawa 252-5210, Japan}
\email{joten.okamoto@nao.ac.jp}
\date{\today}

\begin{abstract}

Fine-structure dynamics in solar prominences holds critical clues to understanding their physical nature of significant space-weather implications. We report evidence of rotational motions of horizontal helical threads in two active-region prominences observed by the \emph{Hinode} and/or \emph{IRIS} satellites at high resolution. In the first event, we found transverse motions of brightening threads at speeds up to 55~km~s$^{-1}$ seen in the plane of the sky. Such motions appeared as sinusoidal space--time trajectories with a typical period of $\sim$390~s, which is consistent with plane-of-sky projections of rotational motions. Phase delays at different locations suggest propagation of twists along the threads at phase speeds of 90--270~km~s$^{-1}$. At least 15 episodes of such motions occurred in two days, none associated with any eruption. For these episodes, the plane-of-sky speed is linearly correlated with the vertical travel distance, suggestive of a constant angular speed. In the second event, we found Doppler velocities of 30--40~km~s$^{-1}$ in opposite directions in the top and bottom portions of the prominence, comparable to the plane-of-sky speed. The moving threads have about twice broader line widths than stationary threads. These observations, when taken together, provide strong evidence for rotations of helical prominence threads, which were likely driven by unwinding twists triggered by magnetic reconnection between twisted prominence magnetic fields and ambient coronal fields. 

\end{abstract}


\section{Introduction}
\label{sect_intro}

Solar prominences are dense, cool plasma residing at elevated heights in the tenuous, hot corona \citep[for recent reviews, see ][]{lab10,mac10}. According to their relative locations with respect to active regions, they are generally classified into three types: active-region, quiescent, and intermediate prominences \citep[e.g., ][]{eng98,mac10,via15}. They have two magnetic-field configurations: normal \citep{kip57} and inverse \citep{kup74}, depending on whether the predominantly horizontal magnetic field threading the prominence material is in the same or opposite direction with respect to the underlying ambient field. There is a correlation between the two configurations and the maximum height of the prominence \citep{ler84}. The inverse configuration is of particular significance, because it can host a helical magnetic flux rope \citep{kar03}, believed to be the progenitor a coronal mass ejection \citep[CME; e.g.,][]{pne83,pri89,van90,rus94,ama99,low05}. Evidence for the inverse configuration or its helical structure has been obtained from polarimetric measurements of the photosphere and chromosphere \citep[e.g.,][]{lit05,lop06,oka09,kuc12} and observations of erupting prominences and associated CMEs \citep[e.g.][]{yur02,mar04,vrs04,gib06,liu15}. 

Although a prominence as a whole can appear macroscopically stationary, it often consists of numerous microscopically dynamic threads.  Fine-structure horizontal threads have been identified in active-region prominences and vertical threads in hedgerow, quiescent prominences \citep{eng76,lin05}. Horizontal threads can travel along magnetic field lines \citep{zir98} and exhibit transverse oscillations as manifestations of Alfv\'enic waves \citep[e.g.,][]{oka07,lin09}. In addition, recent high-resolution observations revealed a variety of small-scale dynamics, such as rising dark plumes and bubbles \citep{ber08,ber10}, upward plasma blob ejections \citep{hil11}, descending knots \citep{cha10}, and horizontally-moving bright blobs \citep{lin12}. Such motions can provide useful diagnostics for various physical quantities, e.g., with a technique called prominence seismology \citep{ter08,sol09,oli09,arr12}. 

Among various prominence fine-structure dynamics, rotational motions are of particular importance, because they are directly relevant to magnetic helicity and free energy, key parameters governing the evolution and ultimate fate a magnetic flux rope. Such motions can be grouped in two categories: flows of mass along largely static, helical magnetic field lines and bodily rotations involving magnetic fields themselves. Category~I includes flows along circular paths (projected onto the plane of the sky) detected in H$\alpha$ images and spectra \citep{ohm69,lig84} and in coronal cavities seen in EUV images \citep{li12,pan13}. Category~II includes structural rotations of the prominence (together with its frozen-in magnetic fields) as a whole about its horizontal spine \citep{wil09} or in part about a vertical axis within a so-called ``tornado" \citep{oro12,su14}, the latter of which was argued to be actually a prominence barb where counter-streaming flows along fine threads \citep{lin03} were mistaken as apparent rotations \citep{pan14}. 

All the prior observations of prominence rotations were obtained with limited spatial and/or temporal resolution and thus left considerable ambiguities such as the debate on the nature of ``tornadoes". High spatio-temporal and spectral resolutions are in critical need to resolve such ambiguities and to provide new insights to the details of the physical processes involved. Here we present such observations of two active-region prominences showing clear signatures of rotational motions of horizontal threads. Both events were observed by the Solar Optical Telescope \citep[SOT;][]{tsu08,shi08,sue08} on board the \emph{Hinode} satellite \citep{kos07} offering the highest spatial resolution of the Sun ever achieved in space. The second event was also observed by the \emph{Interface Region Imaging Spectrograph} \citep[\emph{IRIS};][]{dep14} providing critical Doppler measurements with high spectral resolution. In what follows, we present observations and data analysis results in Section~\ref{sect_obs} and interpretations and discussions in Section~\ref{sect_dis}.

\section{Observations}
\label{sect_obs}

\subsection{Event~1: 2007 February 8--10}
\label{sect_event1}

The \emph{Hinode} satellite observed the west limb of the Sun from 11:00~UT on 2007 February 8 to 10:49~UT on February 10. We obtained 18,889 images of an active region prominence in NOAA AR 10940 with the SOT \ion{Ca}{2} H-line filter (3968 \AA, bandwidth 3 \AA) at a spatial sampling of $0 \farcs 12$. We had a 108\arcsec$\times$108\arcsec\ (1024$\times$1024~pixel$^2$) field of view (FOV), a 0.3~s exposure time, and an 8~s cadence with occasional interruptions because of synoptic observations and data losses. We note that this dataset were also analyzed by \citet{nis08} and \citet{liu09} for helical chromospheric jets and \citet{zha12} for mass flows in the prominence.

\begin{figure}
\epsscale{1}
\plotone{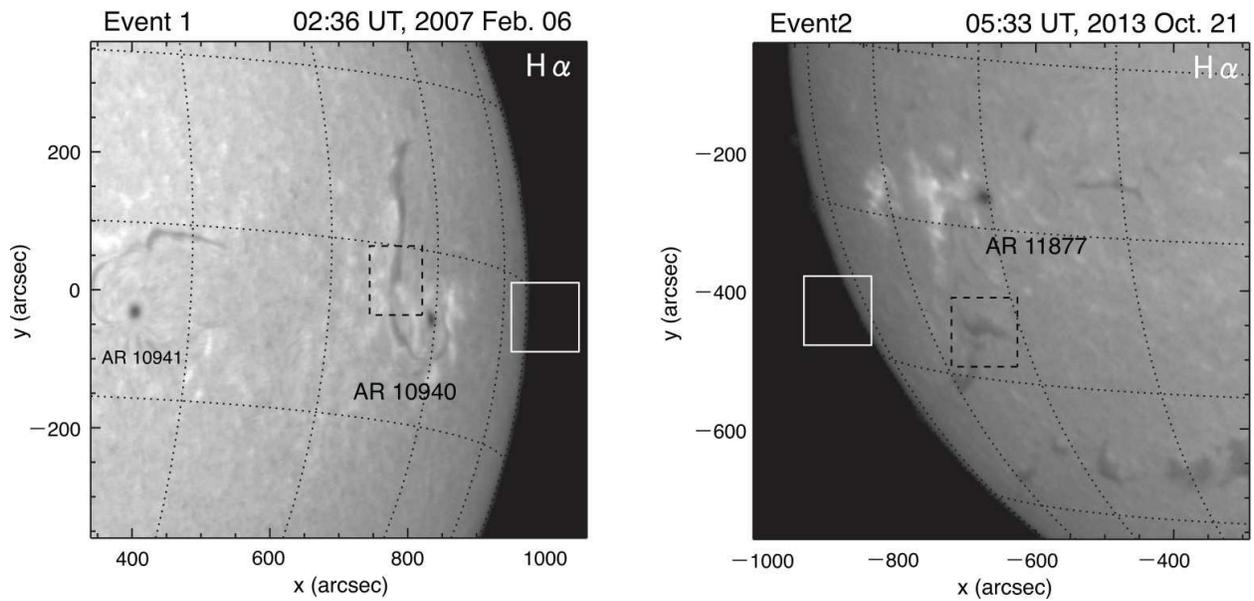}
\caption{
H$\alpha$ images from the Solar Magnetic Activity Research Telescope (SMART) at the Hida Observatory showing the two prominences of interest and their associated active regions, two days before (left) and after (right) the events under study when they appeared at the solar limb. The white boxes indicate the \emph{Hinode}/SOT's FOV at the limb, while the black dashed boxes mark the observing targets at their corresponding disk locations. 
}
\label{fig1}
\end{figure}

\begin{figure}
\epsscale{1}
\plotone{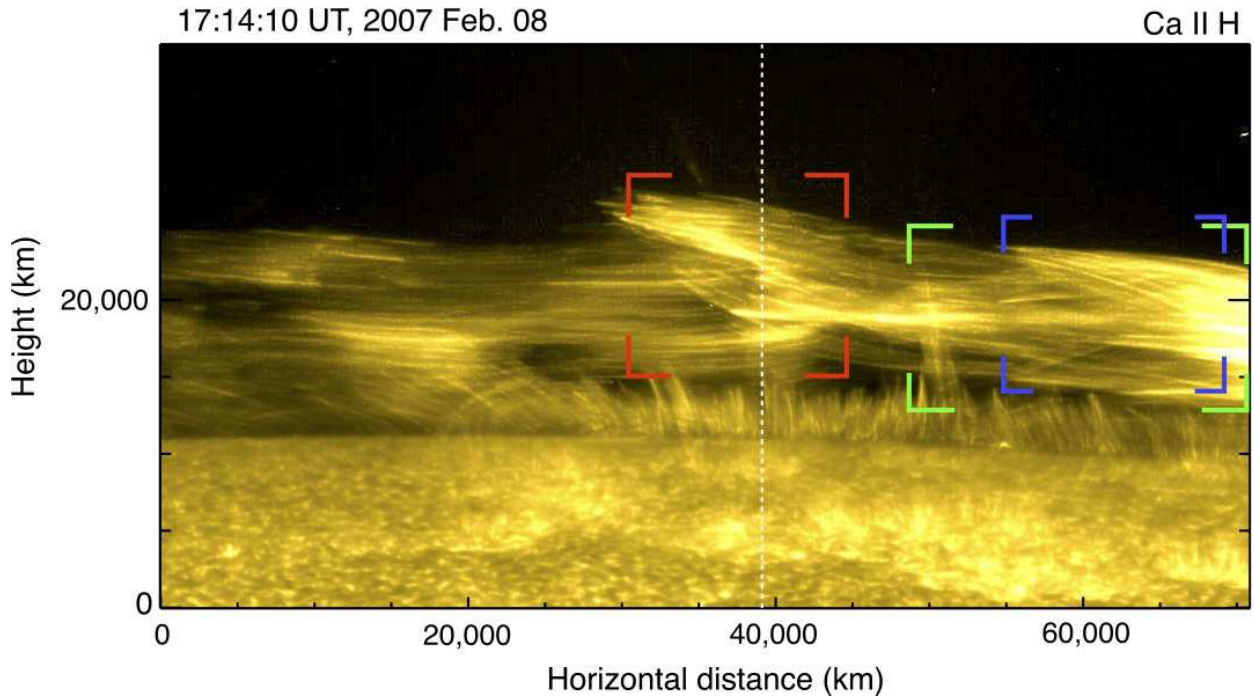}
\caption{
Snapshot of the prominence in Event~1 observed in the \ion{Ca}{2}~H line by \emph{Hinode}/SOT at 17:14:10~UT on 2007 February 8. The solar north is to the left and the west is up. The red brackets indicate the FOV of Figure~\ref{fig3} showing the sudden appearance of bright threads. The blue and green brackets mark the regions for rotational motions shown in Figure~\ref{fig4} and propagating waves shown in Figure~\ref{fig5}, respectively. The white dotted line indicates a vertical cut that was used to make the space-time plot in Figure~\ref{fig6}.
(An animation of this figure is available.)
}
\label{fig2}
\end{figure}

\begin{figure}
\epsscale{1}
\plotone{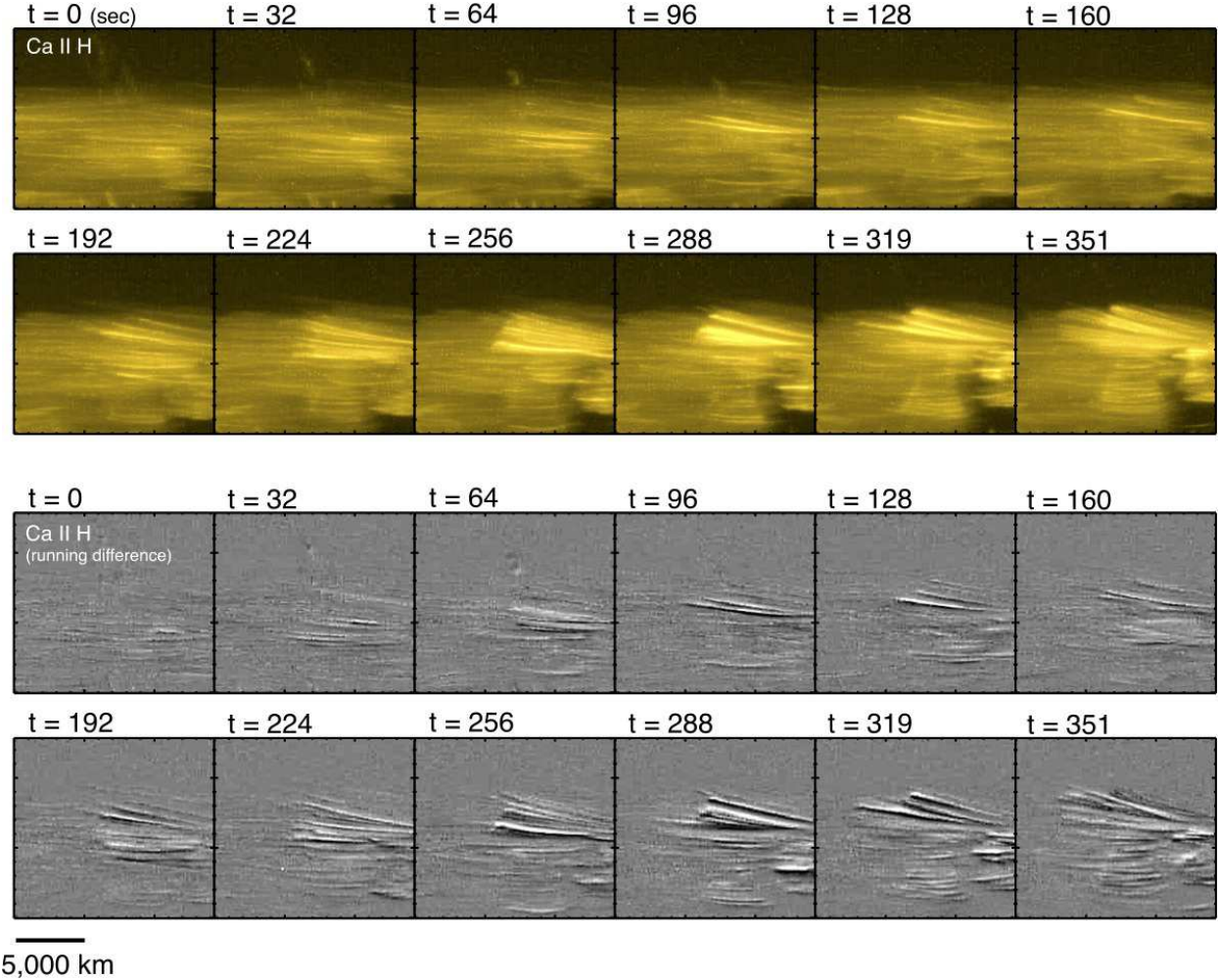}
\caption{
Time-series of \ion{Ca}{2}~H images in the red-boxed region of Figure~\ref{fig2} showing prominence threads undergoing brightening. The two top rows are intensity maps, while the two bottom rows are running differences between consecutive images to emphasize the transversely moving threads that appear as adjacent bright and dark stripes. The reference ($t=0$) for the time stamps (in sec) is 16:58:50~UT on February 8.
}
\label{fig3}
\end{figure}

The prominence was elongated in the north-south direction according to H$\alpha$ images taken two days earlier (see the left panel of Figure~\ref{fig1}). Figure~\ref{fig2} shows a snapshot of \ion{Ca}{2}~H-line observed at 17:14~UT on February 8 (also see Movie 1). This prominence consisted of horizontal fine threads with a typical width of 350~km ($0 \farcs 5$), as is the case with a typical active region prominence \citep[e.g.,][]{oka07}. While the prominence configuration as a whole remained stationary, each individual thread had gentle horizontal flows and small vertical oscillations. At times, a bundle of bright threads suddenly appeared in the prominence and moved transversely upward, with each thread following one another sequentially. The threads were not exactly parallel to the solar limb, but slightly inclined. After reaching their maximum heights, some threads turned around and moved transversely downward or flowed horizontally along presumably magnetic field lines \citep[also discussed in][]{zha12}. An example of this process is shown in Figure~\ref{fig3}.

\begin{figure}
\epsscale{1}
\plotone{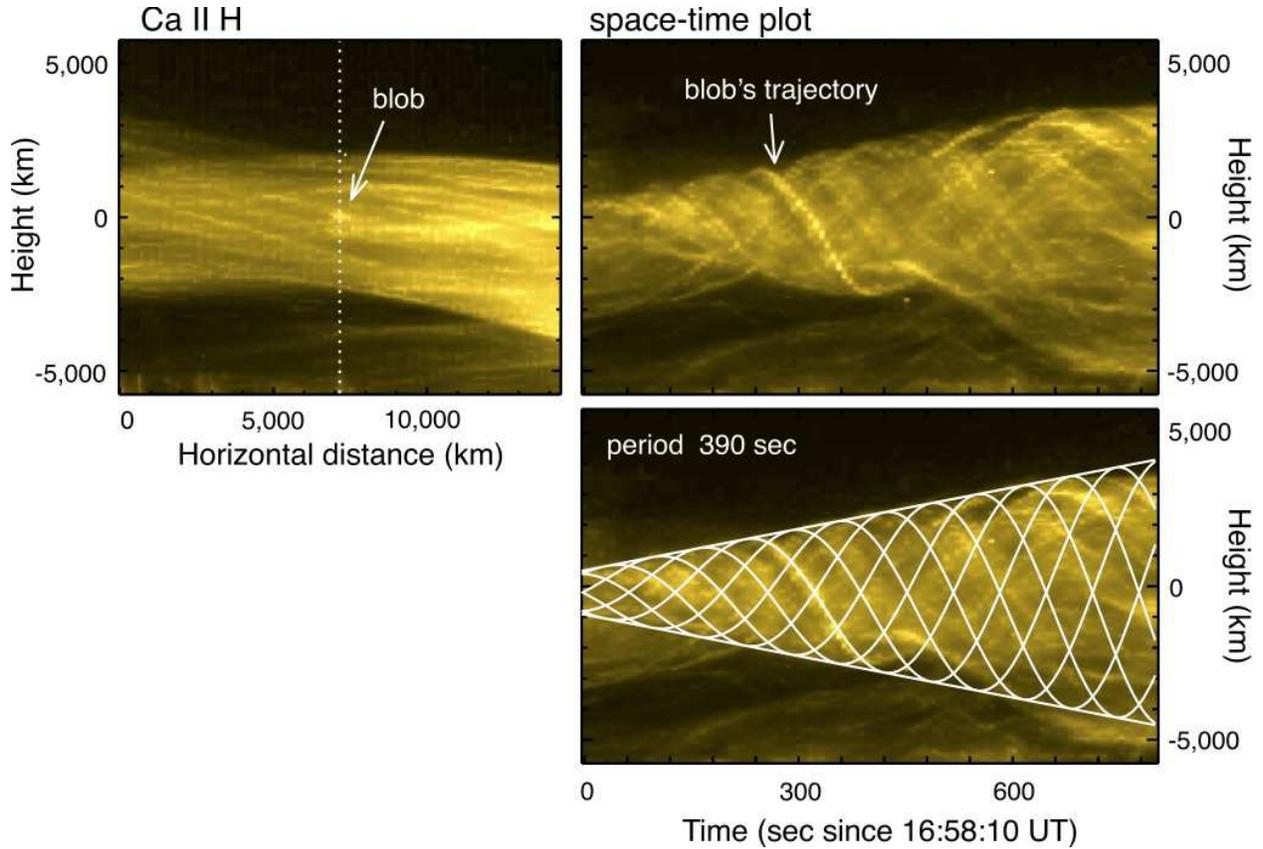}
\caption{
Snapshot (left panel) showing prominence threads in the blue box of Figure~\ref{fig2} and space--time plots (right panels) obtained from a vertical cut (dashed line on the left) located on a bright blob. The two right panels are identical, but the bottom one is overlaid with representative sinusoidal curves that correspond to projected trajectories of assumed rotations (period $\sim$390~s) of a bundle of threads that expand radially at a rate of 4.5~km~s$^{-1}$. The origin of time on the horizontal axis is 16:58:10~UT on February 8.
}
\label{fig4}
\end{figure}

We selected a bright blob within a thread and made a space--time plot using a vertical cut to track its motion (see top panels in Figure~\ref{fig4}). The blob appeared at the top of the prominence and then moved downward. Interestingly, its space--time trajectory has a sinusoidal shape, which is consistent with a line-of-sight projection of rotation about an axis nearly parallel to the plane of the sky. The plane-of-sky speed varied with time and had a maximum of 55~km~s$^{-1}$ near the middle of the prominence and approached zero at the top and bottom edges. The lower-right panel in Figure~\ref{fig4} shows the same space--time plot superimposed with projected trajectories expected from a rotational motion with a period of 390~s and a radius increasing at a rate of 4.5~km~s$^{-1}$. Notably the brightest trajectory near $t=300$~s, corresponding to the selected bright blob, is well fitted by the presentative curve.

\begin{figure}
\epsscale{1}
\plotone{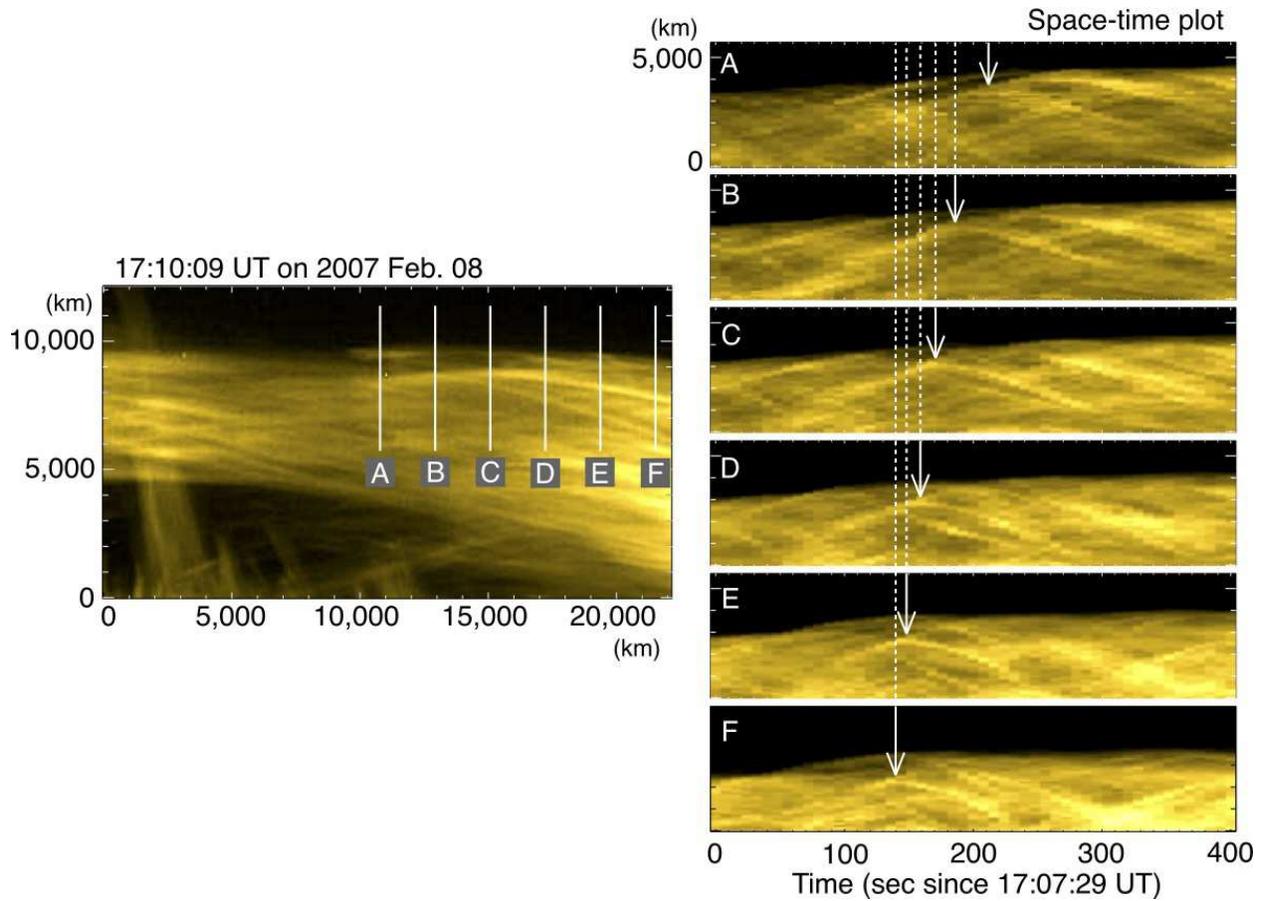}
\caption{
Left: Snapshot image of swaying threads seen in the green box shown in Figure~\ref{fig2}. Right: Space--time plots obtained from vertical cuts~A\,--\,F, evenly spaced in the horizontal direction, as shown on the left. We identified one prominent thread captured by all cuts. Its corresponding peak excursions in space--time plots are delayed in time, indicative of propagation of twists at speeds of 270~km~s$^{-1}$ from cut F to E and 90~km~s$^{-1}$ from cut B to A. The reference time is 17:07:29~UT on February 8.
}
\label{fig5}
\end{figure}

We detected phase delays of sinusoids in space--time plots from different vertical cuts as shown in Figure~\ref{fig5}. If the inferred rotational motion is the case, the phase delays indicate that the threads are not parallel, but inclined to their rotational axis along the prominence spine. In other words, the threads are not aligned along a straight cylinder, but in a helical shape. Specifically, the phase delay from cut~F to A in Figure~\ref{fig5} suggests a propagation of twists from right to left, i.e., in the direction toward the first appearance of bright threads. The time delays and spatial separations between the cuts translate to a propagation speed $v_{\rm phase}$ that drops with distance from 270 to 90~km~s$^{-1}$. Considering the typical period of 390~s noted above, the wavelength or pitch of helix thus falls in the range of 35\,--\,105~Mm. We note a typical value of 20~km~s$^{-1}$ for the maximum transverse speed measured in Figure~\ref{fig5}, which corresponds to the rotational speed $v_{\phi}$. We thus obtain the pitch angle of the helical threads $\theta= \arctan(v_{\phi}/v_{\rm phase})$ in the range of 4$^\circ$\,--\,13$^\circ$.

\begin{figure}
\epsscale{1}
\plotone{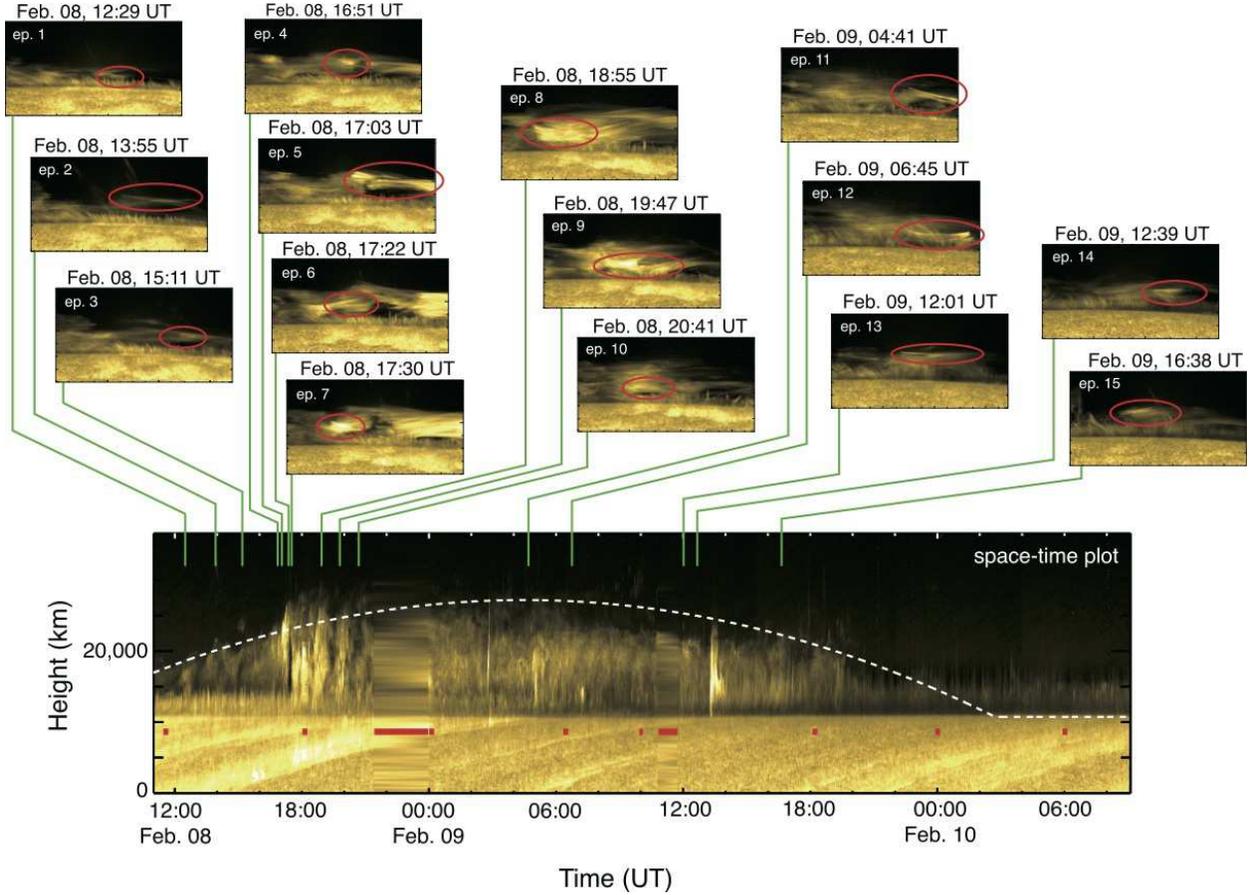}
\caption{
Diagram of prominence evolution over two days for Event~1. The upper panels are snapshots of each episode of recurrent activations. The red ovals mark the appearances of bright threads. The lower panel is a space--time plot obtained from a selected vertical cut along the white dotted line in Figure~\ref{fig2}. The white dashed curve indicates the plane-of-sky projection of the prominence height assumed to be constant at 16,500~km ($\sim$23\arcsec), whose temporal variation is caused by the solar rotation. The red horizontal bars in the lower panel indicate times of occasional interruptions and data loss.
}
\label{fig6}
\end{figure}

Within the two days of observations, appearances of bright thread bundles and subsequent transverse motions repeatedly took place for at least 15 episodes (Figure~\ref{fig6}). None of these episodes was associated with any flare or eruptive phenomenon. Each episode lasted from a few to 20 minutes and the bright threads did not reach beyond the maximum height of the prominence. One episode (\#14) showed only downward motions of bright threads, while the other episodes had both upward and downward motions. From the corresponding space--time plots, we measured the maximum velocity $v_{\rm max}$, vertical travel distance $\Delta y$, and duration of the vertical motions of the bright threads in each episode, which are summarized in Table~\ref{tab1}. Note that these quantities are typical values for each episode, not for individual threads. 

\begin{figure}
\epsscale{0.6}
\plotone{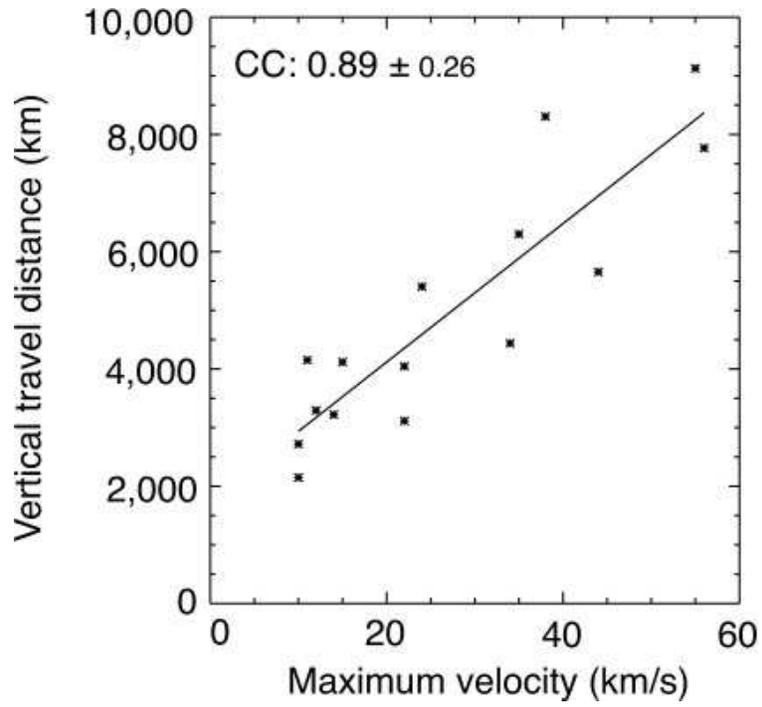}
\caption{
Scatter plot showing the correlation between the maximum velocity and the travel distance of vertical motions of bright threads for Event~1. The solid line represents a linear fit to the data, which has a linear correlation coefficient of $0.89 \pm 0.26$.
}
\label{fig7}
\end{figure}

Interestingly, as shown in Figure~\ref{fig7}, the maximum velocity and vertical travel distance are strongly correlated with a linear correlation coefficient of $0.89\pm 0.26$. This is again consistent with the inferred plane-of-sky projections of prominence rotations, if $v_{\rm max}$ and $\Delta y$ correspond to the rotational speed and diameter, respectively, and thus implies a constant angular velocity or period. Under this assumption, we estimate the rotation period to be $P= \pi \Delta y/ v_{\rm max}$, which has a statistical mean of 370~s for all 15 episodes. This value is close to the period of 390~s estimated for one of the episodes using sinusoidal fits to space--time trajectories, as shown in Figure~\ref{fig4}. That approach could not be applied to the rest of the episodes, because of lack of well-defined crests and troughs in those trajectories mainly due to line-of-sight overlaps of numerous threads. We note in passing that the linear fit in Figure~\ref{fig7} does not go through the origin, possibly due to a systematic overestimate of the vertical travel distance.

\subsection{Event~2: 2013 October 19}
\label{sect_event2}

We conducted coordinated \emph{Hinode}--\emph{IRIS} observations of an active region prominence in NOAA AR~11877 (right panel of Figure~\ref{fig1} and also see Movie 2) at the southeast limb on 2013 October 19. \emph{Hinode}/SOT took 585 images with the \ion{Ca}{2}~H line filter at a cadence of 8~s from 09:09 to 10:27~UT. The FOV was 108\arcsec$\times$108\arcsec\ (1024$\times$1024 pixel$^2$) and the exposure time was 1.2~s. \emph{IRIS} performed 360 sets of four-step sparse raster scans from 09:00 to 11:00~UT. The cadence was 20~s between slit locations with spatial and spectral resolutions of $0 \farcs 33$--$0 \farcs 40$ (240\,--\,290~km on the Sun) and 3~km~s$^{-1}$, respectively. The slit covered 175\arcsec\ length and the exposure time was 4~s. 

\begin{figure}
\epsscale{1}
\plotone{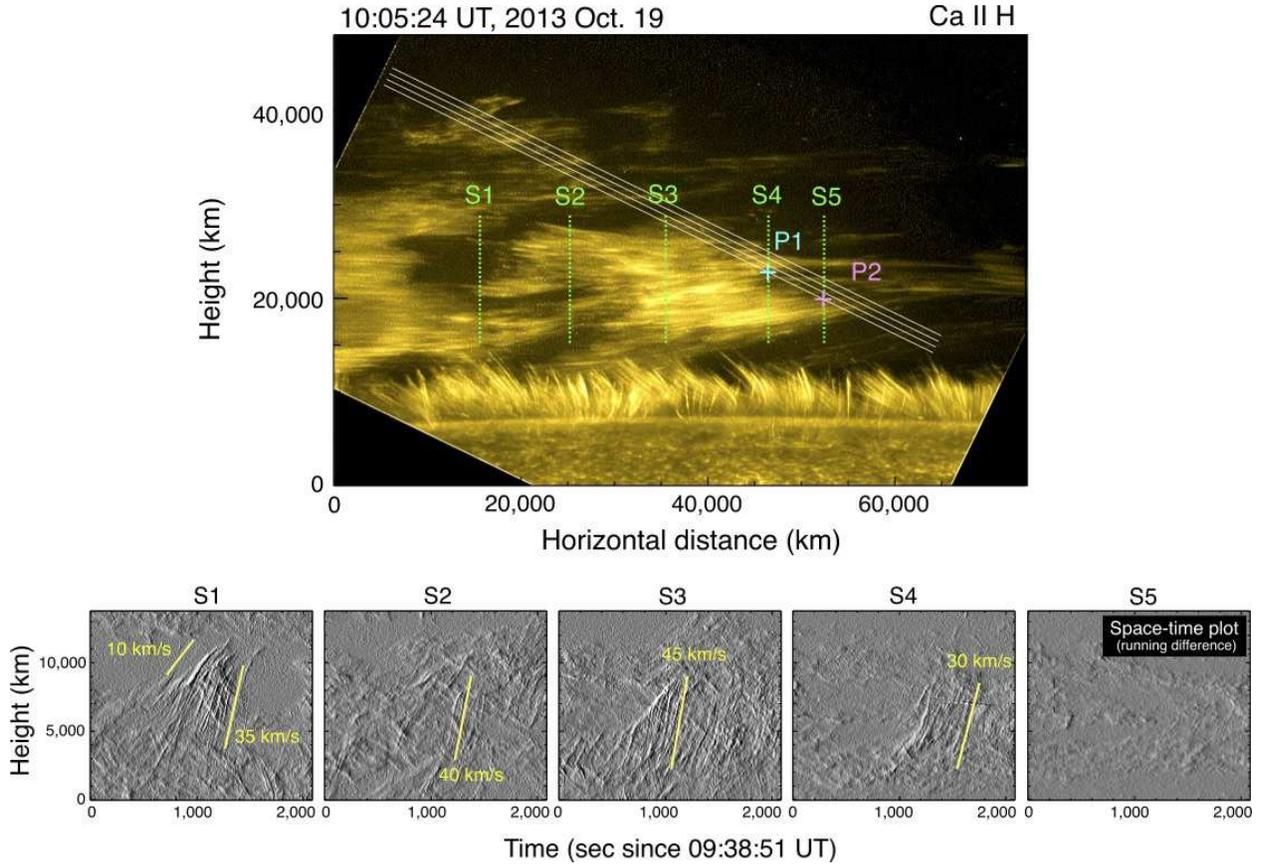}
\caption{
Top: Snapshot of the prominence in Event~2 observed in the \ion{Ca}{2}~H line by \emph{Hinode}/SOT at 10:05:24~UT on 2013 October 19. The four inclined white lines indicate the \emph{IRIS} slit locations. The green dashed lines (S1--S5) cover the locations of the bundles of appearing bright threads. Two locations at P1 and P2 are used to show the spectra of bright threads in Figures~\ref{fig9} and \ref{fig10}. The FOV is rotated to have the solar limb oriented horizontally. Bottom: Space--time plots of running difference images obtained from vertical cuts marked by S1--S5 in the top panel.
(An animation of this figure is available.)
}
\label{fig8}
\end{figure}

Figure~\ref{fig8} shows the observed prominence. We note that this is the same prominence analyzed by \citet{oka15} and \citet{ant15}, who found evidence of resonant absorption of transverse oscillations and associated heating. They focused on the higher portion of the prominence where minimum line-of-sight overlap allowed for spatially resolving fine threads and small-scale dynamics. Here we focused on the prominence's lower portion, which shows evidence of rotational motions with amplitudes on the order of 10,000~km, about 10 times those of the transverse oscillations in the higher portion of the prominence reported by \citet{oka15}. First, we examined the plane-of-sky motions using SOT data. Similar to Event~1, while the prominence was stationary as a whole, sudden appearances of bundles of bright threads took place. The bottom panels in Figure~\ref{fig8} show space--time plots of running-difference images obtained from vertical cuts~S1--S5 shown as green dashed lines in the top panel. The bright threads moved upward sequentially at typical speeds of 30\,--\,45~km~s$^{-1}$. The top edge of the bundle, best seen within cuts~S1 and S3, also shifted upward at a speed of $\sim$10~km~s$^{-1}$. We note that only a small portion of the prominence threads near their right end was covered by the \emph{IRIS} slits. The lowest slit position intersects cuts~S4 and S5 at locations marked by plus signs P1 and P2, respectively. As clearly seen in the online movie, the transverse motions of the threads and ensuing horizontal flows mainly occurred to the left of cut~S5 whose space--time plot shows no clear signature of such motions.

Next, we analyzed the spectral data taken by \emph{IRIS}. We focused on the strong \ion{Mg}{2}~k line that has sufficient signal to noise ratio required for spectral fits. In general, the \ion{Mg}{2}~k line profile in on-disk observations has a central reversal in the emission peak within an even broader absorption component, which are produced by a complex radiative transfer process \citep{lee13a,lee13b,per13}. In off-limb observations, as long as there is no spatial overlap of different features along the line of sight, the profile is simple with a single, Gaussian-shaped peak indicative of optically thin emission \citep[e.g., ][]{sch14,liu15,via16}. This is indeed the case for most parts of our observed prominence. Let us examine the spectra at the locations of the bright threads. Figure~\ref{fig9} shows time-series of the spectra before and after the bright threads passed through the positions at the cyan and pink plus signs (P1 and P2) in Figure~\ref{fig8}. At P1, the spectrum in the beginning had a single Gaussian profile with a negligible Doppler velocity ($\leq$4~km~s$^{-1}$) that corresponds to the rest component. As time progresses, while the rest component remained without significant changes, an enhancement in the \emph{blue wing} appeared with a Doppler velocity up to $-35$~km~s$^{-1}$. This blueshifted component represents a new feature appearing in front of or behind (along the line of sight) the pre-existing rest component. Likewise, the spectrum at P2 also showed double components, but with a dominant \emph{red-wing} enhancement. Noting that P1 and P2 are respectively located in the top and bottom portions of the bundle of threads, these opposite Doppler shifts indicate a clockwise rotation\footnote{Note that there is a possibility that these opposite Doppler shifts are indications of counter-streaming flows, commonly seen in prominence threads \citep[e.g., ][]{lin03}. However, this possibility is very small in our case, given the two observational facts that (1) the opposite Doppler shifts are detected right at the top and bottom edges of the transverse motions of the thread bundle, and (2) the Doppler and transverse velocities have similar magnitudes of $\sim$40~km~s$^{-1}$. Therefore, these facts naturally fit in the picture of prominence rotation, but it would be very coincidental for a counter-streaming scenario to be applicable here.} (when viewed from left to right in Figure~\ref{fig8}) of the prominence about an horizontal axis within/near the plane of the sky. To satisfy the continuity condition, the threads seen transversely moving upward must be located on the back side of the rotating structure.

Figure~\ref{fig10} shows history of the Doppler velocity and the line width at positions P1 and P2 derived from double or single Gaussian fits (see examples in Figure~\ref{fig9}). The bright prominence threads exhibited large Doppler velocities up to 30--40~km~s$^{-1}$ of dominant blueshifts at P1 and redshifts at P2. Both components had broad line widths of $\sim$20~km~s$^{-1}$, about twice those of the corresponding rest components. We note that the dominant blue components at P1 appeared about 14~min earlier and had slightly greater Doppler velocities than the red ones at P2. This is because the \emph{IRIS} slit was inclined to the threads and thus it sampled different parts of the horizontal extent of the prominence, with a lag in arrival time of horizontal flows at the slit locations.


\section{Summary and Discussion}
\label{sect_dis}

We have reported two events in which bright bundles of horizontal prominence threads appeared and exhibited transverse motions in the vertical direction. In the first event (see Section~\ref{sect_event1}), we found sinusoidal trajectories of bright threads in space--time plots, which are consistent with plane-of-sky projections of rotational motions. Clear phase delays suggest that those threads are not straight, but of helical shapes. Further and more decisive evidence of rotation was found in the second event (Section~\ref{sect_event2}) exhibiting similar transverse motions in the sky plane at speeds up to 45~km~s$^{-1}$, which were accompanied by Doppler velocities of comparable magnitudes up to 30\,--\,40~km~s$^{-1}$ in opposite directions near the top and bottom edges of the prominence. The combination of consistent plane-of-sky and line-of-sight motions observed here allows us to conclude that these events involved bodily rotations of helical prominence threads. 

Although rotations have been observed in prominences \citep[e.g.,][]{wil09} and helical jets \citep[e.g.,][]{liu09}, our observations were obtained with the highest spatial and spectral resolution ever achieved together in space. The rotations of helical threads reported here fall into the second category of prominence rotations mentioned in Section~\ref{sect_intro}, i.e., structural rotations rather than mass flows along field lines (Category~I). The novelty of these observations lies in the fact that such rotations involve \emph{certain portions of horizontal threads}, which differs from previously reported Category~II rotations of \emph{an entire prominence} \citep[e.g.,][]{wil09} or of the debatable ``prominence tornado" about a \emph{vertical rotational axis} \citep[e.g.,][]{su14}. We provide below an interpretation of the observed phenomenon and discuss its implications.

\subsection{Interpretation and Basic Scenarios}

This phenomenon occurred episodically in otherwise stationary prominences, with no indication of solar flares, jets, or other catastrophic processes. This means that the integrity of the overall magnetic geometry must remain largely unchanged through these episodes. Whatever process is, changes to the magnetic field must be minor and gentle, and the involved energy release is small, insufficient to cause an eruption. With these inferences, we propose two possible scenarios, schematically shown in Figure~\ref{fig11}, to explain the observations. In both scenarios, magnetic reconnection plays an important role in transferring twists and magnetic helicity from highly twisted fields to less twisted or untwisted fields and in triggering the unwinding rotational motions of helical prominence threads.

In Scenario~A (Figure~\ref{fig11}, left), when highly twisted helical flux emerges into the atmosphere and reaches the height of prominence threads of straight or less twisted magnetic field, magnetic reconnection between the two flux systems can occur. Magnetic helicity and twists can be transferred to the prominence field so as to cause it to rotate. The propagation of twists can also explain the observed phase delays along the prominence threads for projected sinusoidal transverse motions.

In Scenario~B (Figure~\ref{fig11}, right), the helical prominence magnetic field is assumed to be more twisted than the ambient field, e.g., of the neighboring coronal loops. Magnetic reconnection between them would cause transfer of twists and helicity in an opposite direction, i.e., away from the prominence, compared with Scenario~A. As such, the helical prominence threads would undergo unwinding rotational motions and relaxation until a new equilibrium state is reached with twists evenly distributed along the entire, newly reconnected field line.

Our observations suggest that Scenario~B is more likely. In the case of the helical flux emergence (Scenario~A), the effects of rotations would propagate upward from the lower atmosphere, where we found no such signature. In addition, flux emergence, because of its usually large free magnetic energy content, could cause more dramatic response such as jets \citep{kur87}, as detected in the same \emph{Hinode}/SOT dataset for Event~1 \citep{nis08,liu09}. In both events here, the change of the magnetic configuration appeared gentle. It is thus more likely the case that reconnection occurred between twisted prominence fields and ambient coronal loops. Moreover, the observed twist propagation is consistent with this scenario, in which the unwinding motion traveled from the twisted part of the prominence to the less-twisted part where the brightening threads first appeared, suggestive of heating or energy release caused by magnetic reconnection there. Note that the rate or amount of energy release is likely not high enough to cause significant heating at coronal temperatures of a sufficiently large volume to be detected by AIA, because we found no clear signature of brightening in AIA passbands (which also have a spatial resolution of $1 \farcs 5$, much smaller than \emph{Hinode} or \emph{IRIS}).

\subsection{Discussion on Helical Motions of Prominence Threads}

\begin{figure}
\epsscale{0.7}
\plotone{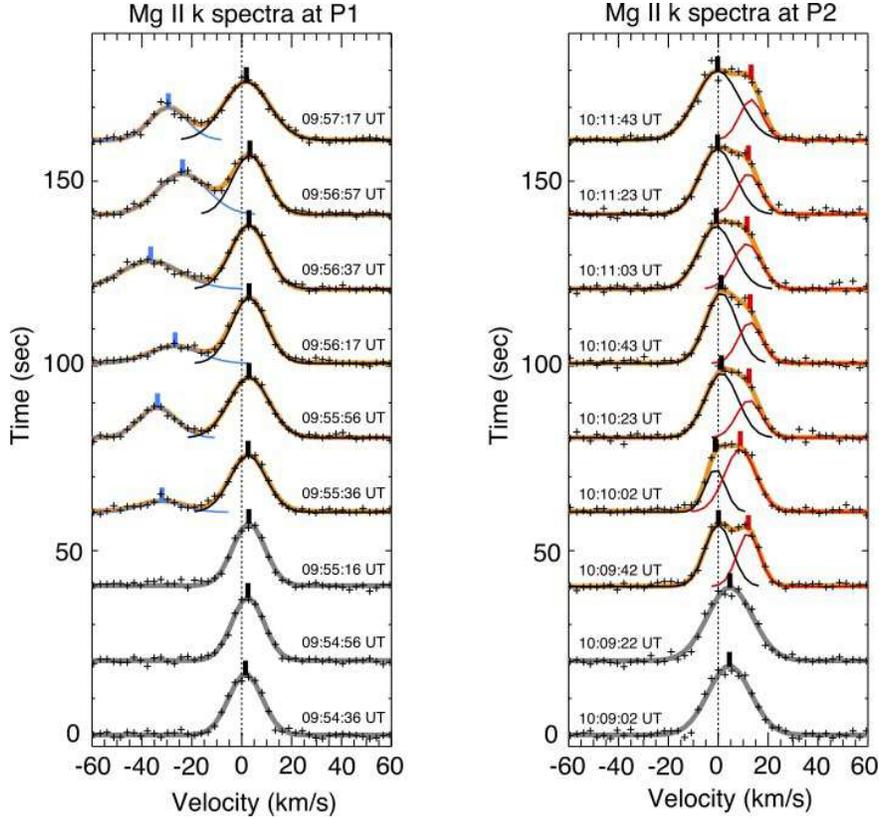}
\caption{
Time series of \ion{Mg}{2}~k spectra at the positions of the two plus signs P1 and P2 shown in Figure~\ref{fig8}. Black plus signs here present the original data. Thick gray and orange curves are single and double Gaussian fits, respectively. For double Gaussian fits, the thin black curve is the rest component, while the thin blue/red curve is the blue-/red-shifted component. The thick black and blue/red bars indicate the central wavelengths of the corresponding Gaussian components.
}
\label{fig9}
\end{figure}

\begin{figure}
\epsscale{1}
\plotone{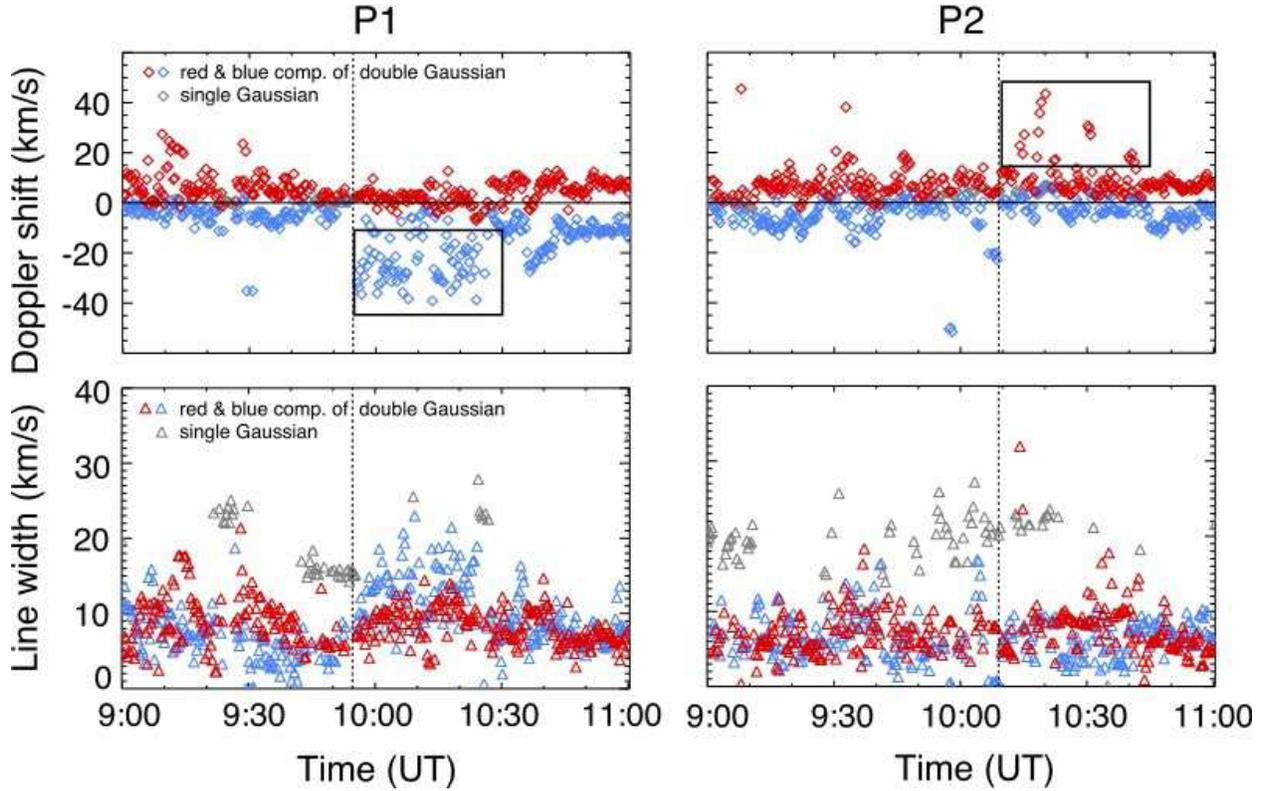}
\caption{
Temporal evolution of the Doppler shift and line width at positions P1 and P2 shown in Figure~\ref{fig8}. Red and blue symbols represent red and blue components derived from double Gaussian fits, respectively. Gray symbols are from single Gaussian fits in cases of failure of double Gaussian fits. The vertical dotted lines indicate the start times of the spectra shown in Figure~\ref{fig9}, which exhibit dominant blueshifts at P1 and redshifts at P2, as marked by the black boxes.
}
\label{fig10}
\end{figure}

\begin{figure}
\epsscale{1}
\plotone{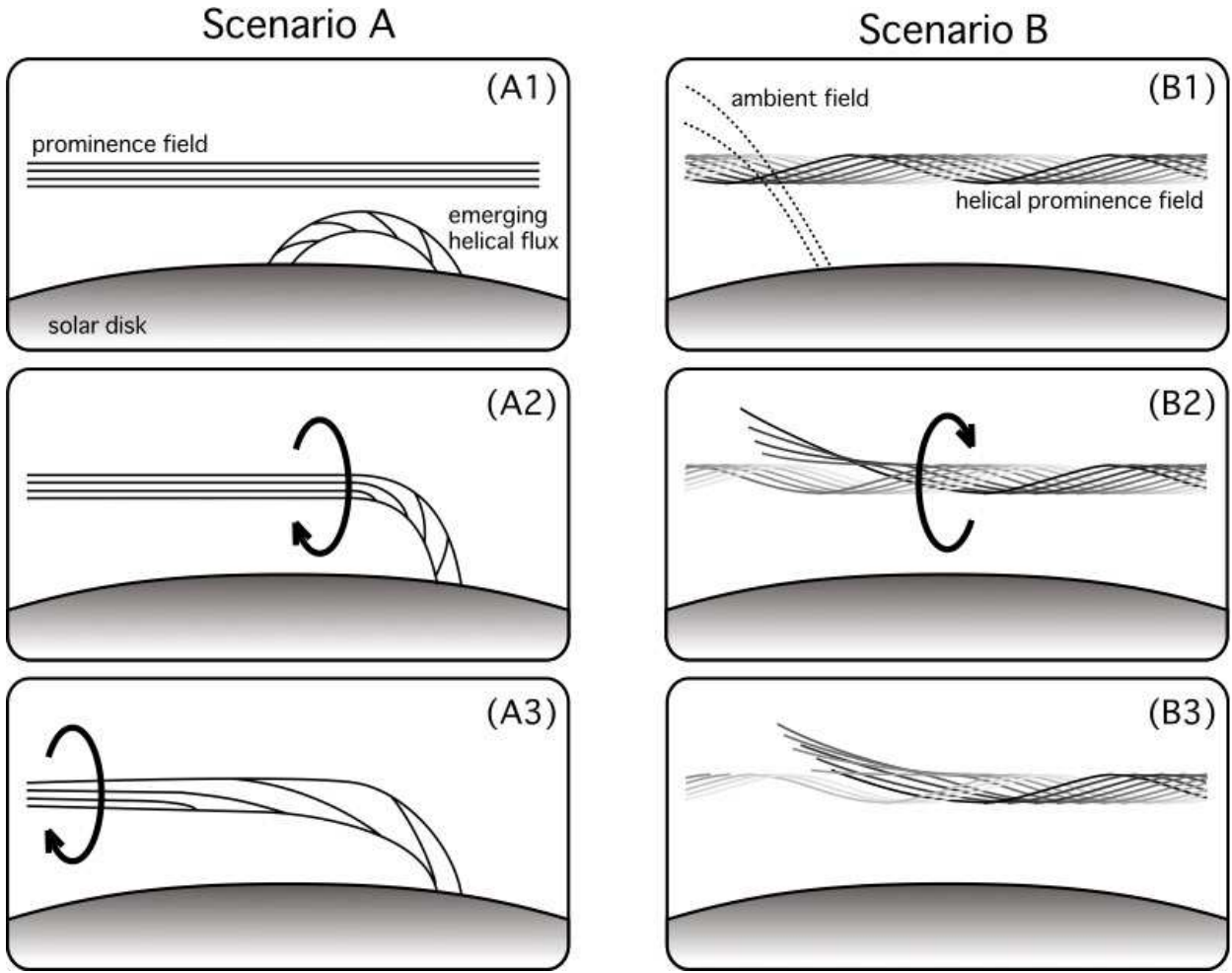}
\caption{
Schematic of two possible scenarios that can explain the observed prominence rotation. Both scenarios involve magnetic reconnection between twisted and untwisted field lines and transfer of twists and magnetic helicity. Left: Emerging helical flux reconnects with the prominence field and triggers its rotation. Right: Helical prominence field reconnects with the ambient field and the ensuing unwinding relaxation causes the prominence to spin. The numbered labels in each scenario indicate the temporal development.
}
\label{fig11}
\end{figure}

Our observed prominence activations were not associated with drastic activities such as flares. In contrast, with on-disk observations by the \emph{Hinode} EUV Imaging Spectrometer (EIS), \citet{wil09} detected a prominence with a somewhat smaller rotational speed of 20~km~s$^{-1}$ before a flare occurred. In their case, the entire prominence rotated synchronously like a cylinder. They suggested that the rotation was caused by some change in the surrounding magnetic environment, such as small-scale flux emergence \citep{moo92} which is consistent with our Scenario~A, or by some form of the so-called break-out process \citep{ant99,aul00}. The change of magnetic configuration affected the entire prominence and thus led to its synchronous rotation. In our case, the prominence, at least in Event~2, was not rotating entirely, because we detected both the rotational and stationary components simultaneously (Figure~\ref{fig9}). Therefore, it is more likely that reconnection took place only in some part of the prominence magnetic field, as shown in Figure~\ref{fig11} for Scenario~B.

Measurements of magnetic helicity or twists can provide critical clues to the generation of magnetic fields either by convective dynamo action or differential rotation \citep[see e.g.,][]{pev95}. This is one of the most important issues in prominence research in regard to the mechanism of prominence formation. However, in both events, the pitch angle of the helical prominence threads and thus of their frozen-in magnetic-field lines is very small, on the order of $10^\circ$ or smaller. This introduces large uncertainty or ambiguity in determining the sign of the pitch angle with respect to the rotational axis. Therefore, we could not infer the handedness of the helical field lines here, unlike in other case \citep[e.g.,][]{liu15} with large pitch angles combined with simultaneous plane-of-sky and Doppler velocity measurements.

The {\it Hinode} and {\it IRIS} observations presented here, at a level of clarity not previously achieved, have offered new insights to rotational motions of fine-structure prominence threads. Additional observations in even greater detail, including temperature measurements, will soon be provided by the Atacama Large Millimeter/submillimeter Array (ALMA) in its observing Cycle~4 starting 2016 \citep[see][]{wed16}. We anticipate to utilize the combination of such unprecedented observations to further probe the physics of prominence activations and related phenomena.

\acknowledgements 

\emph{Hinode} is a Japanese mission developed and launched by ISAS/JAXA, with NAOJ as domestic partner and NASA and STFC (UK) as international partners. It is operated by these agencies in cooperation with ESA and NSC (Norway). \emph{IRIS} is a NASA small explorer mission developed and operated by LMSAL with mission operations executed at NASA Ames Research center and major contributions to downlink communications funded by ESA and the Norwegian Space Centre. T.J.O. thanks Ippon-kakou-kai for their encouragement and was supported by JSPS KAKENHI Grant Number 25800120/16K17663 (PI: T.J.O.) and 25220703 (PI: S.~Tsuneta). W.L. is supported by IRIS Guest Investigator grant NNX15AR15G and NASA contract NNG09FA40C (IRIS). This work was partly carried out on the Solar Data Analysis System operated by the Astronomy Data Center in cooperation with the Hinode Science Center of NAOJ.



\begin{table*}[htbp]
\begin{center}
\caption{
Characteristics of detected episodes of transverse motions in the vertical direction of horizontal prominence threads in Event~1.}
\begin{tabular}{lcccc}
\\
\hline
Episode & Maximum velocity & Duration & Vertical travel distance  \\
     & (km~s$^{-1}$)    & (s)  & (km)  \\
\hline
 1 & 22 & 270 & 3120 \\
 2 & 10 & 540 & 2150 \\
 3 & 11 & 510 & 4150 \\
 4 & 34 & 480 & 4440 \\
 5 & 55 & 1220 & 9130 \\
 6 & 56 & 490 & 7770 \\
 7 & 24 & 400 & 5410 \\
 8 & 22 & 620 & 4050 \\
 9 & 35 & 1220 & 6300 \\
10 & 10 & 1130 & 2720 \\
11 & 38 & 990 & 8310 \\
12 & 44 & 960 & 5660 \\
13 & 12 & 400 & 3300 \\
14 & 14 & 840 & 3200 \\
15 & 15 & 990 & 4120 \\
\hline
Range  & 10--56 & 270--1220 & 2150--9130 \\
Mean   & 27     & 737       & 4922 \\
Median & 22     & 620       & 4150 \\
\hline 
\label{tab1}
\end{tabular}
\end{center}
\end{table*}

\end{document}